# Cross section measurement of $e^+e^- \to \pi^+\pi^-\psi(3686)$ from $\sqrt{s} = 4.0076$ GeV to 4.6984 GeV

M. Ablikim,[1] M. N. Achasov,[10,b] P. Adlarson,[67] S. Ahmed,[15] M. Albrecht,[4] R. Aliberti,[28] A. Amoroso,[66a,66c] M. R. An,[32] Q. An,[63,49] X. H. Bai,[57] Y. Bai,[48] O. Bakina,[29] R. Baldini Ferroli,[23a] I. Balossino,[24a] Y. Ban,[38,i] K. Begzsuren,[26] N. Berger,[28] M. Bertani,[23a] D. Bettoni,[24a] F. Bianchi,[66a,66c] J. Bloms,[60] A. Bortone,[66a,66c] I. Boyko,[29] R. A. Briere,[5] H. Cai,[68] X. Cai,[1,49] A. Calcaterra,[23a] G. F. Cao,[1,54] N. Cao,[1,54] S. A. Cetin,[53a] J. F. Chang,[1,49] W. L. Chang,[1,54] G. Chelkov,[29,a] D. Y. Chen,[6] G. Chen,[1] H. S. Chen,[1,54] M. L. Chen,[1,49] S. J. Chen,[35] X. R. Chen,[25] Y. B. Chen,[1,49] Z. J. Chen,[20,j] W. S. Cheng,[66c] G. Cibinetto,[24a] F. Cossio,[66c] X. F. Cui,[36] H. L. Dai,[1,49] X. C. Dai,[1,54] A. Dbeyssi,[15] R. E. de Boer,[4] D. Dedovich,[29] Z. Y. Deng,[1] A. Denig,[28] I. Denysenko,[29] M. Destefanis,[66a,66c] F. De Mori,[66a,66c] Y. Ding,[33] C. Dong,[36] J. Dong,[1,49] L. Y. Dong,[1,54] M. Y. Dong,[1,49,54] X. Dong,[68] S. X. Du,[71] Y. L. Fan,[68] J. Fang,[1,49] S. S. Fang,[1,54] Y. Fang,[1] R. Farinelli,[24a] L. Fava,[66b,66c] F. Feldbauer,[4] G. Felici,[23a] C. Q. Feng,[63,49] J. H. Feng,[50] M. Fritsch,[4] C. D. Fu,[1] Y. Gao,[64] Y. Gao,[63,49] Y. Gao,[38,i] Y. G. Gao,[6] I. Garzia,[24a,24b] P. T. Ge,[68] C. Geng,[50] E. M. Gersabeck,[58] A. Gilman,[61] K. Goetzen,[11] L. Gong,[33] W. X. Gong,[1,49] W. Gradl,[28] M. Greco,[66a,66c] L. M. Gu,[35] M. H. Gu,[1,49] S. Gu,[2] Y. T. Gu,[13] C. Y. Guan,[1,54] A. Q. Guo,[22] L. B. Guo,[34] R. P. Guo,[40] Y. P. Guo,[9,g] A. Guskov,[29,a] T. T. Han,[41] W. Y. Han,[32] X. Q. Hao,[16] F. A. Harris,[56] K. L. He,[1,54] F. H. Heinsius,[4] C. H. Heinz,[28] T. Held,[4] Y. K. Heng,[1,49,54] C. Herold,[51] M. Himmelreich,[11,e] T. Holtmann,[4] G. Y. Hou,[1,54] Y. R. Hou,[54] Z. L. Hou,[1] H. M. Hu,[1,54] J. F. Hu,[47,k] T. Hu,[1,49,54] Y. Hu,[1] G. S. Huang,[63,49] L. Q. Huang,[64] X. T. Huang,[41] Y. P. Huang,[1] Z. Huang,[38,i] T. Hussain,[65] N. Hüsken,[22,28] W. Ikegami Andersson,[67] W. Imoehl,[22] M. Irshad,[63,49] S. Jaeger,[4] S. Janchiv,[26] Q. Ji,[1] Q. P. Ji,[16] X. B. Ji,[1,54] X. L. Ji,[1,49] Y. Y. Ji,[41] H. B. Jiang,[41] X. S. Jiang,[1,49,54] J. B. Jiao,[41] Z. Jiao,[18] S. Jin,[35] Y. Jin,[57] M. Q. Jing,[1,54] T. Johansson,[67] N. Kalantar-Nayestanaki,[55] X. S. Kang,[33] R. Kappert,[55] M. Kavatsyuk,[55] B. C. Ke,[43,1] I. K. Keshk,[4] A. Khoukaz,[60] P. Kiese,[28] R. Kiuchi,[1] R. Kliemt,[11] L. Koch,[30] O. B. Kolcu,[53a,d] B. Kopf,[4] M. Kuemmel,[4] M. Kuessner,[4] A. Kupsc,[67] M. G. Kurth,[1,54] W. Kühn,[30] J. J. Lane,[58] J. S. Lange,[30] P. Larin,[15] A. Lavania,[21] L. Lavezzi,[66a,66c] Z. H. Lei,[63,49] H. Leithoff,[28] M. Lellmann,[28] T. Lenz,[28] C. Li,[39] C. H. Li,[32] Cheng Li,[63,49] D. M. Li,[71] F. Li,[1,49] G. Li,[1] H. Li,[63,49] H. Li,[43] H. B. Li,[1,54] H. J. Li,[16] J. L. Li,[41] J. Q. Li,[4] J. S. Li,[50] Ke Li,[1] L. K. Li,[1] Lei Li,[3] P. R. Li,[31,l,m] S. Y. Li,[52] W. D. Li,[1,54] W. G. Li,[1] X. H. Li,[63,49] X. L. Li,[41] Xiaoyu Li,[1,54] Z. Y. Li,[50] H. Liang,[63,49] H. Liang,[1,54] H. Liang,[27] Y. F. Liang,[45] Y. T. Liang,[25] G. R. Liao,[12] L. Z. Liao,[1,54] J. Libby,[21] C. X. Lin,[50] B. J. Liu,[1] C. X. Liu,[1] D. Liu,[15,63] F. H. Liu,[44] Fang Liu,[1] Feng Liu,[6] H. B. Liu,[13] H. M. Liu,[1,54] Huanhuan Liu,[1] Huihui Liu,[17] J. B. Liu,[63,49] J. L. Liu,[64] J. Y. Liu,[1,54] K. Liu,[1] K. Y. Liu,[33] L. Liu,[63,49] M. H. Liu,[9,g] P. L. Liu,[1] Q. Liu,[68] Q. Liu,[54] S. B. Liu,[63,49] Shuai Liu,[46] T. Liu,[1,54] W. M. Liu,[63,49] X. Liu,[31,l,m] Y. Liu,[31,l,m] Y. B. Liu,[36] Z. A. Liu,[1,49,54] Z. Q. Liu,[41] X. C. Lou,[1,49,54] F. X. Lu,[50] H. J. Lu,[18] J. D. Lu,[1,54] J. G. Lu,[1,49] X. L. Lu,[1] Y. Lu,[1] Y. P. Lu,[1,49] C. L. Luo,[34] M. X. Luo,[70] P. W. Luo,[50] T. Luo,[9,g] X. L. Luo,[1,49] X. R. Lyu,[54] F. C. Ma,[33] H. L. Ma,[1] L. L. Ma,[41] M. M. Ma,[1,54] Q. M. Ma,[1] R. Q. Ma,[1,54] R. T. Ma,[54] X. X. Ma,[1,54] X. Y. Ma,[1,49] F. E. Maas,[15] M. Maggiora,[66a,66c] S. Maldaner,[4] S. Malde,[61] Q. A. Malik,[65] A. Mangoni,[23b] Y. J. Mao,[38,i] Z. P. Mao,[1] S. Marcello,[66a,66c] Z. X. Meng,[57] J. G. Messchendorp,[55] G. Mezzadri,[24a] T. J. Min,[35] R. E. Mitchell,[22] X. H. Mo,[1,49,54] Y. J. Mo,[6] N. Yu. Muchnoi,[10,b] H. Muramatsu,[59] S. Nakhoul,[11,e] Y. Nefedov,[29] F. Nerling,[11,e] I. B. Nikolaev,[10,b] Z. Ning,[1,49] S. Nisar,[8,h] S. L. Olsen,[54] Q. Ouyang,[1,49,54] S. Pacetti,[23b,23c] X. Pan,[9,g] Y. Pan,[58] A. Pathak,[1] A. Pathak,[27] P. Patteri,[23a] M. Pelizaeus,[4] H. P. Peng,[63,49] K. Peters,[11,e] J. Pettersson,[67] J. L. Ping,[34] R. G. Ping,[1,54] S. Pogodin,[29] R. Poling,[59] V. Prasad,[63,49] H. Qi,[63,49] H. R. Qi,[52] K. H. Qi,[25] M. Qi,[35] T. Y. Qi,[9] S. Qian,[1,49] W. B. Qian,[54] Z. Qian,[50] C. F. Qiao,[54] L. Q. Qin,[12] X. P. Qin,[9] X. S. Qin,[41] Z. H. Qin,[1,49] J. F. Qiu,[1] S. Q. Qu,[36] K. H. Rashid,[65] K. Ravindran,[21] C. F. Redmer,[28] A. Rivetti,[66c] V. Rodin,[55] M. Rolo,[66c] G. Rong,[1,54] Ch. Rosner,[15] M. Rump,[60] H. S. Sang,[63] A. Sarantsev,[29,c] Y. Schelhaas,[28] C. Schnier,[4] K. Schoenning,[67] M. Scodeggio,[24a,24b] D. C. Shan,[46] W. Shan,[19] X. Y. Shan,[63,49] J. F. Shangguan,[46] M. Shao,[63,49] C. P. Shen,[9] H. F. Shen,[1,54] P. X. Shen,[36] X. Y. Shen,[1,54] H. C. Shi,[63,49] R. S. Shi,[1,54] X. Shi,[1,49] X. D. Shi,[63,49] J. J. Song,[41] W. M. Song,[27,1] Y. X. Song,[38,i] S. Sosio,[66a,66c] S. Spataro,[66a,66c] K. X. Su,[68] P. P. Su,[46] F. F. Sui,[41] G. X. Sun,[1] H. K. Sun,[1] J. F. Sun,[16] L. Sun,[68] S. S. Sun,[1,54] T. Sun,[1,54] W. Y. Sun,[27] W. Y. Sun,[34] X. Sun,[20,j] Y. J. Sun,[63,49] Y. K. Sun,[63,49] Y. Z. Sun,[1] Z. T. Sun,[1] Y. H. Tan,[68] Y. X. Tan,[63,49] C. J. Tang,[45] G. Y. Tang,[1] J. Tang,[50] J. X. Teng,[63,49] V. Thoren,[67] W. H. Tian,[43] Y. T. Tian,[25] I. Uman,[53b] B. Wang,[1] C. W. Wang,[35] D. Y. Wang,[38,i] H. J. Wang,[31,l,m] H. P. Wang,[1,54] K. Wang,[1,49] L. L. Wang,[1] M. Wang,[41] M. Z. Wang,[38,i] Meng Wang,[1,54] W. Wang,[50] W. H. Wang,[68] W. P. Wang,[63,49] X. Wang,[38,i] X. F. Wang,[31,l,m] X. L. Wang,[9,g] Y. Wang,[63,49] Y. Wang,[50] Y. D. Wang,[37] Y. F. Wang,[1,49,54] Y. Q. Wang,[1] Y. Y. Wang,[31,l,m] Z. Wang,[1,49] Z. Y. Wang,[1] Ziyi Wang,[54] Zongyuan Wang,[1,54] D. H. Wei,[12] F. Weidner,[60] S. P. Wen,[1] D. J. White,[58] U. Wiedner,[4] G. Wilkinson,[61] M. Wolke,[67] L. Wollenberg,[4] J. F. Wu,[1,54] L. H. Wu,[1] L. J. Wu,[1,54] X. Wu,[9,g] Z. Wu,[1,49] L. Xia,[63,49] H. Xiao,[9,g] S. Y. Xiao,[1] Z. J. Xiao,[34] X. H. Xie,[38,i] Y. G. Xie,[1,49] Y. H. Xie,[6] T. Y. Xing,[1,54] G. F. Xu,[1] Q. J. Xu,[14] W. Xu,[1,54] X. P. Xu,[46] Y. C. Xu,[54] F. Yan,[9,g] L. Yan,[9,g] W. B. Yan,[63,49] W. C. Yan,[71] Xu Yan,[46] H. J. Yang,[42,f] H. X. Yang,[1] L. Yang,[43] S. L. Yang,[54] Y. X. Yang,[12] Yifan Yang,[1,54] Zhi Yang,[25] M. Ye,[1,49] M. H. Ye,[7] J. H. Yin,[1] Z. Y. You,[50] B. X. Yu,[1,49,54] C. X. Yu,[36] G. Yu,[1,54] J. S. Yu,[20,j] T. Yu,[64] C. Z. Yuan,[1,54]

L. Yuan,[2] X. Q. Yuan,[38,i] Y. Yuan,[1] Z. Y. Yuan,[50] C. X. Yue,[32] A. A. Zafar,[65] X. Zeng Zeng,[6] Y. Zeng,[20,j] A. Q. Zhang,[1] B. X. Zhang,[1] Guangyi Zhang,[16] H. Zhang,[63] H. H. Zhang,[50] H. H. Zhang,[27] H. Y. Zhang,[1,49] J. J. Zhang,[43] J. L. Zhang,[69] J. Q. Zhang,[34] J. W. Zhang,[1,49,54] J. Y. Zhang,[1] J. Z. Zhang,[1,54] Jianyu Zhang,[1,54] Jiawei Zhang,[1,54] L. M. Zhang,[52] L. Q. Zhang,[50] Lei Zhang,[35] S. Zhang,[50] S. F. Zhang,[35] Shulei Zhang,[20,j] X. D. Zhang,[37] X. Y. Zhang,[41] Y. Zhang,[61] Y. T. Zhang,[71] Y. H. Zhang,[1,49] Yan Zhang,[63,49] Yao Zhang,[1] Z. H. Zhang,[6] Z. Y. Zhang,[68] G. Zhao,[1] J. Zhao,[32] J. Y. Zhao,[1,54] J. Z. Zhao,[1,49] Lei Zhao,[63,49] Ling Zhao,[1] M. G. Zhao,[36] Q. Zhao,[1] S. J. Zhao,[71] Y. B. Zhao,[1,49] Y. X. Zhao,[25] Z. G. Zhao,[63,49] A. Zhemchugov,[29,a] B. Zheng,[64] J. P. Zheng,[1,49] Y. Zheng,[38,i] Y. H. Zheng,[54] B. Zhong,[34] C. Zhong,[64] L. P. Zhou,[1,54] Q. Zhou,[1,54] X. Zhou,[68] X. K. Zhou,[54] X. R. Zhou,[63,49] X. Y. Zhou,[32] A. N. Zhu,[1,54] J. Zhu,[36] K. Zhu,[1] K. J. Zhu,[1,49,54] S. H. Zhu,[62] T. J. Zhu,[69] W. J. Zhu,[36] W. J. Zhu,[9,g] Y. C. Zhu,[63,49] Z. A. Zhu,[1,54] B. S. Zou,[1] and J. H. Zou[1]

(BESIII Collaboration)

[1]Institute of High Energy Physics, Beijing 100049, People's Republic of China
[2]Beihang University, Beijing 100191, People's Republic of China
[3]Beijing Institute of Petrochemical Technology, Beijing 102617, People's Republic of China
[4]Bochum Ruhr-University, D-44780 Bochum, Germany
[5]Carnegie Mellon University, Pittsburgh, Pennsylvania 15213, USA
[6]Central China Normal University, Wuhan 430079, People's Republic of China
[7]China Center of Advanced Science and Technology, Beijing 100190, People's Republic of China
[8]COMSATS University Islamabad, Lahore Campus, Defence Road, Off Raiwind Road, 54000 Lahore, Pakistan
[9]Fudan University, Shanghai 200443, People's Republic of China
[10]G.I. Budker Institute of Nuclear Physics SB RAS (BINP), Novosibirsk 630090, Russia
[11]GSI Helmholtzcentre for Heavy Ion Research GmbH, D-64291 Darmstadt, Germany
[12]Guangxi Normal University, Guilin 541004, People's Republic of China
[13]Guangxi University, Nanning 530004, People's Republic of China
[14]Hangzhou Normal University, Hangzhou 310036, People's Republic of China
[15]Helmholtz Institute Mainz, Staudinger Weg 18, D-55099 Mainz, Germany
[16]Henan Normal University, Xinxiang 453007, People's Republic of China
[17]Henan University of Science and Technology, Luoyang 471003, People's Republic of China
[18]Huangshan College, Huangshan 245000, People's Republic of China
[19]Hunan Normal University, Changsha 410081, People's Republic of China
[20]Hunan University, Changsha 410082, People's Republic of China
[21]Indian Institute of Technology Madras, Chennai 600036, India
[22]Indiana University, Bloomington, Indiana 47405, USA
[23a]INFN Laboratori Nazionali di Frascati, I-00044, Frascati, Italy
[23b]INFN Sezione di Perugia, I-06100, Perugia, Italy
[23c]University of Perugia, I-06100, Perugia, Italy
[24a]INFN Sezione di Ferrara, I-44122, Ferrara, Italy
[24b]University of Ferrara, I-44122, Ferrara, Italy
[25]Institute of Modern Physics, Lanzhou 730000, People's Republic of China
[26]Institute of Physics and Technology, Peace Ave. 54B, Ulaanbaatar 13330, Mongolia
[27]Jilin University, Changchun 130012, People's Republic of China
[28]Johannes Gutenberg University of Mainz, Johann-Joachim-Becher-Weg 45, D-55099 Mainz, Germany
[29]Joint Institute for Nuclear Research, 141980 Dubna, Moscow region, Russia
[30]Justus-Liebig-Universitaet Giessen, II. Physikalisches Institut, Heinrich-Buff-Ring 16, D-35392 Giessen, Germany
[31]Lanzhou University, Lanzhou 730000, People's Republic of China
[32]Liaoning Normal University, Dalian 116029, People's Republic of China
[33]Liaoning University, Shenyang 110036, People's Republic of China
[34]Nanjing Normal University, Nanjing 210023, People's Republic of China
[35]Nanjing University, Nanjing 210093, People's Republic of China
[36]Nankai University, Tianjin 300071, People's Republic of China
[37]North China Electric Power University, Beijing 102206, People's Republic of China
[38]Peking University, Beijing 100871, People's Republic of China
[39]Qufu Normal University, Qufu 273165, People's Republic of China
[40]Shandong Normal University, Jinan 250014, People's Republic of China
[41]Shandong University, Jinan 250100, People's Republic of China
[42]Shanghai Jiao Tong University, Shanghai 200240, People's Republic of China

[43]*Shanxi Normal University, Linfen 041004, People's Republic of China*
[44]*Shanxi University, Taiyuan 030006, People's Republic of China*
[45]*Sichuan University, Chengdu 610064, People's Republic of China*
[46]*Soochow University, Suzhou 215006, People's Republic of China*
[47]*South China Normal University, Guangzhou 510006, People's Republic of China*
[48]*Southeast University, Nanjing 211100, People's Republic of China*
[49]*State Key Laboratory of Particle Detection and Electronics, Beijing 100049, Hefei 230026, People's Republic of China*
[50]*Sun Yat-Sen University, Guangzhou 510275, People's Republic of China*
[51]*Suranaree University of Technology, University Avenue 111, Nakhon Ratchasima 30000, Thailand*
[52]*Tsinghua University, Beijing 100084, People's Republic of China*
[53a]*Turkish Accelerator Center Particle Factory Group, Istanbul Bilgi University, HEP Res. Cent., 34060 Eyup, Istanbul, Turkey*
[53b]*Near East University, Nicosia, North Cyprus, Mersin 10, Turkey*
[54]*University of Chinese Academy of Sciences, Beijing 100049, People's Republic of China*
[55]*University of Groningen, NL-9747 AA Groningen, The Netherlands*
[56]*University of Hawaii, Honolulu, Hawaii 96822, USA*
[57]*University of Jinan, Jinan 250022, People's Republic of China*
[58]*University of Manchester, Oxford Road, Manchester, M13 9PL, United Kingdom*
[59]*University of Minnesota, Minneapolis, Minnesota 55455, USA*
[60]*University of Muenster, Wilhelm-Klemm-Str. 9, 48149 Muenster, Germany*
[61]*University of Oxford, Keble Rd, Oxford, OX13RH, United Kingdom*
[62]*University of Science and Technology Liaoning, Anshan 114051, People's Republic of China*
[63]*University of Science and Technology of China, Hefei 230026, People's Republic of China*
[64]*University of South China, Hengyang 421001, People's Republic of China*
[65]*University of the Punjab, Lahore-54590, Pakistan*
[66a]*University of Turin and INFN, University of Turin, I-10125, Turin, Italy*
[66b]*University of Eastern Piedmont, I-15121, Alessandria, Italy*
[66c]*INFN, I-10125, Turin, Italy*
[67]*Uppsala University, Box 516, SE-75120 Uppsala, Sweden*
[68]*Wuhan University, Wuhan 430072, People's Republic of China*
[69]*Xinyang Normal University, Xinyang 464000, People's Republic of China*
[70]*Zhejiang University, Hangzhou 310027, People's Republic of China*
[71]*Zhengzhou University, Zhengzhou 450001, People's Republic of China*

Using data samples with a total integrated luminosity of 20.1 $\mathrm{fb}^{-1}$ collected by the BESIII detector operating at the BEPCII collider, the cross section of the process $e^+e^- \to \pi^+\pi^-\psi(3686)$ is measured at center-of-mass energies between 4.0076 and 4.6984 GeV. The measured cross section is consistent with previous results, and with much improved precision. A fit to the measured energy-dependent cross section, which includes three Breit-Wigner functions and a nonresonant contribution, confirms the existence of the charmonium-like states $Y(4220)$, $Y(4390)$, and $Y(4660)$. This is the first observation of the $Y(4660)$ at the BESIII experiment.

[a]Also at the Moscow Institute of Physics and Technology, Moscow 141700, Russia.
[b]Also at the Novosibirsk State University, Novosibirsk, 630090, Russia.
[c]Also at the NRC "Kurchatov Institute", PNPI, 188300, Gatchina, Russia.
[d]Currently at Istanbul Arel University, 34295 Istanbul, Turkey.
[e]Also at Goethe University Frankfurt, 60323 Frankfurt am Main, Germany.
[f]Also at Key Laboratory for Particle Physics, Astrophysics and Cosmology, Ministry of Education; Shanghai Key Laboratory for Particle Physics and Cosmology; Institute of Nuclear and Particle Physics, Shanghai 200240, People's Republic of China.
[g]Also at Key Laboratory of Nuclear Physics and Ion-beam Application (MOE) and Institute of Modern Physics, Fudan University, Shanghai 200443, People's Republic of China.
[h]Also at Harvard University, Department of Physics, Cambridge, Massachusetts 02138, USA.
[i]Also at State Key Laboratory of Nuclear Physics and Technology, Peking University, Beijing 100871, People's Republic of China.
[j]Also at School of Physics and Electronics, Hunan University, Changsha 410082, China.
[k]Also at Guangdong Provincial Key Laboratory of Nuclear Science, Institute of Quantum Matter, South China Normal University, Guangzhou 510006, China.
[l]Also at Frontiers Science Center for Rare Isotopes, Lanzhou University, Lanzhou 730000, People's Republic of China.
[m]Also at Lanzhou Center for Theoretical Physics, Lanzhou University, Lanzhou 730000, People's Republic of China.

## I. INTRODUCTION

Over the past several decades, a series of charmonium-like states with $J^{PC} = 1^{--}$, referred to as the $Y$ states, have been discovered and confirmed by numerous experiments. However, the internal structure of the $Y$ states remains controversial. Possible interpretations include hybrid mesons, meson molecules, hadrocharmonium, and tetraquarks, but none of them has so far proved conclusive [1]. Therefore, comprehensive studies of production and decay patterns, as well as the measurement of resonance parameters from different experiments, are desirable to provide information that will help probe the nature of the $Y$ states.

Dedicated studies of $Y$ states were initially triggered by the discovery of the $Y(4260)$ in $e^+e^- \to \gamma_{\rm ISR}\pi^+\pi^- J/\psi$ using the initial-state-radiation (ISR) approach by the *BABAR* experiment [2], which was subsequently confirmed by CLEO [3,4] and Belle [5,6]. The latest results from the BESIII experiment on the channel $e^+e^- \to \pi^+\pi^- J/\psi$ reveal with unprecedented precision that the structure previously identified as the $Y(4260)$ consists of two structures referred to as $Y(4220)$ and $Y(4320)$ [7]. This observation is also confirmed in the process $e^+e^- \to \pi^0\pi^0 J/\psi$ [8]. Furthermore, similar structures to the $Y(4220)$ also appear in the processes $e^+e^- \to \omega\chi_{c0}$ [9], $\pi^+\pi^- h_c$ [10], $\pi^+ D^0 D^{*-}$ [11], and $\eta J/\psi$ [12]. However, the parameters of the $Y(4220)$ observed in different processes are different. Further studies both theoretically and experimentally are still needed to understand the internal structure of the $Y(4220)$.

Analogously, studies of the process $e^+e^- \to \gamma_{\rm ISR}\pi^+\pi^-\psi(3686)$ have been performed at the Belle and *BABAR* experiments, where two $Y$ states, namely the $Y(4360)$ and $Y(4660)$ [13–15] were observed and confirmed. BESIII measured the cross sections of the process $e^+e^- \to \pi^+\pi^-\psi(3686)$ with much improved precision at center-of-mass (c.m.) energies from 4.0076 to 4.5995 GeV [16]. The existence of the $Y(4360)$ was confirmed, but with a mass close to 4.39 GeV$/c^2$, so we refer to it here as the $Y(4390)$. Furthermore, the $Y(4220)$ was reported for the first time in the final state $\pi^+\pi^-\psi(3686)$, but the $Y(4660)$ was not studied due to the lack of data in the corresponding energy region. It is worth noting that a $Y(4390)$ with close mass is also reported in the processes $e^+e^- \to \pi^+\pi^- h_c$ [10], and $\eta J/\psi$ [12] by BESIII. The differences in parameters of the $Y$ states of different experiments require more studies to understand, and the $e^+e^- \to \pi^+\pi^-\psi(3686)$ process is an appropriate channel to provide more information on the $Y$ states.

In this paper, an analysis of $e^+e^- \to \pi^+\pi^-\psi(3686)$ is presented using an analysis approach similar to that of Ref. [16], with old data sets used in Ref. [16] and new data sets collected by BESIII since 2017. Additional data samples between 4.23 and 4.36 GeV and new data samples between 4.60 and 4.70 GeV are used to study the properties of the $Y$ states, especially the $Y(4660)$. The c.m. energy and the corresponding luminosity [8,17,18] of each of the samples used in this analysis can be found in Table I. Following Ref. [16], the $\psi(3686)$ is reconstructed via its charged decay mode $\psi(3686) \to \pi^+\pi^- J/\psi$ (Mode I) and the neutral decay mode $\psi(3686) \to \text{neutrals} + J/\psi$ (Mode II), where "neutrals" refers to $\pi^0$, $\pi^0\pi^0$ or $\eta$ from $\psi(3686)$ and $\gamma\gamma$ from the cascade decay $\psi(3686) \to \gamma\chi_{cJ} \to \gamma\gamma J/\psi$. Only the neutral decays of $\pi^0$ and $\eta$ are considered in

TABLE I. The Born cross section $\sigma^B$ at individual c.m. energies. The subscript "cha" or "neu" denotes the charged mode (Mode I) or the neutral mode (Mode II), respectively. The uncertainties on the signal yields and the Born cross sections for individual decay modes are statistical only. The first uncertainties for the combined Born cross sections are statistical and the second are systematic.

| $E_{\rm cms}$ (GeV) | $\mathcal{L}_{\rm int}({\rm pb}^{-1})$ | $N_{\rm cha}$ | $\epsilon_{\rm cha}(\%)$ | $N_{\rm neu}$ | $\epsilon_{\rm neu}(\%)$ | $(1+\delta)$ | $(\frac{1}{\vert 1-\Pi\vert^2})$ | $\sigma^B_{\rm cha}$ (pb) | $\sigma^B_{\rm neu}$ (pb) | $\sigma^B$ (pb) |
|---|---|---|---|---|---|---|---|---|---|---|
| 4.0076 | 482.0 | $1.0^{+1.4}_{-0.7}$ | 30.63 | $5.0^{+3.2}_{-2.4}$ | 8.64 | 0.654 | 1.044 | $0.2^{+0.3}_{-0.2}$ | $5.7^{+3.6}_{-2.7}$ | $0.4^{+0.5}_{-0.3} \pm 0.0$ |
| 4.0854 | 52.9 | $4.0^{+2.4}_{-1.7}$ | 42.34 | $1.4^{+1.8}_{-1.1}$ | 28.35 | 0.761 | 1.051 | $5.4^{+3.2}_{-2.3}$ | $3.7^{+5.0}_{-3.0}$ | $5.0^{+2.5}_{-1.9} \pm 0.2$ |
| 4.1285 | 401.5 | $48.9 \pm 7.2$ | 42.95 | $13.1 \pm 4.9$ | 32.61 | 0.778 | 1.052 | $8.4 \pm 1.2$ | $4.0 \pm 1.5$ | $7.1 \pm 1.0 \pm 0.3$ |
| 4.1574 | 408.7 | $60.0 \pm 8.3$ | 43.49 | $34.9 \pm 7.2$ | 34.37 | 0.785 | 1.053 | $9.9 \pm 1.4$ | $9.8 \pm 2.0$ | $9.6 \pm 1.1 \pm 0.4$ |
| 4.1784 | 3194.5 | $489.9 \pm 23.0$ | 43.81 | $294.2 \pm 21.0$ | 35.59 | 0.788 | 1.054 | $10.2 \pm 0.5$ | $10.2 \pm 0.7$ | $10.2 \pm 0.4 \pm 0.5$ |
| 4.1888 | 570.0 | $84.1 \pm 9.5$ | 43.73 | $56.1 \pm 9.2$ | 35.73 | 0.788 | 1.056 | $9.8 \pm 1.1$ | $10.8 \pm 1.8$ | $10.1 \pm 0.9 \pm 0.5$ |
| 4.1989 | 526.0 | $117.4 \pm 11.2$ | 44.35 | $58.2 \pm 9.8$ | 34.02 | 0.788 | 1.056 | $14.6 \pm 1.4$ | $12.8 \pm 2.1$ | $14.1 \pm 1.2 \pm 0.6$ |
| 4.2091 | 572.1 | $114.6 \pm 11.2$ | 43.95 | $71.6 \pm 10.4$ | 34.23 | 0.786 | 1.057 | $13.3 \pm 1.3$ | $14.4 \pm 2.1$ | $13.6 \pm 1.1 \pm 0.6$ |

*(Table continued)*

TABLE I. *(Continued)*

| $E_{\rm cms}$ (GeV) | $\mathcal{L}_{\rm int}({\rm pb}^{-1})$ | $N_{\rm cha}$ | $\epsilon_{\rm cha}(\%)$ | $N_{\rm neu}$ | $\epsilon_{\rm neu}(\%)$ | $(1+\delta)$ | $(\frac{1}{\vert 1-\Pi\vert^2})$ | $\sigma^B_{\rm cha}$ (pb) | $\sigma^B_{\rm neu}$ (pb) | $\sigma^B$ (pb) |
|---|---|---|---|---|---|---|---|---|---|---|
| 4.2185 | 569.2 | $153.9 \pm 12.9$ | 43.82 | $107.4 \pm 12.4$ | 35.41 | 0.782 | 1.056 | $18.1 \pm 1.5$ | $21.1 \pm 2.4$ | $19.0 \pm 1.3 \pm 0.9$ |
| 4.2263 | 1100.9 | $347.2 \pm 19.3$ | 45.30 | $186.9 \pm 17.4$ | 35.80 | 0.778 | 1.056 | $20.5 \pm 1.1$ | $18.9 \pm 1.8$ | $20.0 \pm 1.0 \pm 1.0$ |
| 4.2357 | 530.3 | $150.7 \pm 12.8$ | 43.60 | $83.4 \pm 11.4$ | 35.10 | 0.823 | 1.056 | $18.1 \pm 1.5$ | $16.8 \pm 2.3$ | $17.7 \pm 1.3 \pm 0.9$ |
| 4.2436 | 594.0 | $162.1 \pm 13.5$ | 43.45 | $86.6 \pm 12.1$ | 35.27 | 0.840 | 1.056 | $17.1 \pm 1.4$ | $15.2 \pm 2.1$ | $16.6 \pm 1.2 \pm 0.8$ |
| 4.2580 | 828.4 | $249.0 \pm 16.2$ | 44.15 | $106.8 \pm 13.7$ | 32.91 | 0.812 | 1.054 | $19.2 \pm 1.3$ | $15.0 \pm 1.9$ | $18.1 \pm 1.0 \pm 0.9$ |
| 4.2668 | 531.1 | $168.4 \pm 13.4$ | 43.98 | $88.5 \pm 12.1$ | 27.95 | 0.805 | 1.053 | $20.6 \pm 1.6$ | $23.0 \pm 3.1$ | $21.1 \pm 1.5 \pm 1.0$ |
| 4.2777 | 175.7 | $73.6 \pm 8.7$ | 42.94 | $42.2 \pm 8.1$ | 24.01 | 0.799 | 1.053 | $28.0 \pm 3.3$ | $38.8 \pm 7.4$ | $30.3 \pm 3.1 \pm 1.5$ |
| 4.2879 | 502.4 | $202.4 \pm 14.6$ | 42.80 | $114.7 \pm 13.3$ | 27.20 | 0.796 | 1.053 | $27.1 \pm 2.0$ | $32.7 \pm 3.8$ | $28.5 \pm 1.8 \pm 1.4$ |
| 4.3079 | 45.1 | $17.6 \pm 4.3$ | 45.08 | $19.4 \pm 5.0$ | 38.17 | 0.792 | 1.052 | $25.1 \pm 6.1$ | $44.1 \pm 11.5$ | $31.7 \pm 5.6 \pm 1.4$ |
| 4.3121 | 501.2 | $279.1 \pm 17.1$ | 43.78 | $196.1 \pm 16.2$ | 37.97 | 0.792 | 1.052 | $36.9 \pm 2.3$ | $40.4 \pm 3.3$ | $38.1 \pm 1.9 \pm 1.6$ |
| 4.3374 | 505.0 | $392.7 \pm 20.6$ | 44.63 | $287.7 \pm 19.3$ | 41.04 | 0.792 | 1.051 | $50.6 \pm 2.7$ | $54.5 \pm 3.7$ | $52.0 \pm 2.2 \pm 2.2$ |
| 4.3583 | 543.9 | $499.5 \pm 23.0$ | 45.75 | $338.9 \pm 21.3$ | 42.19 | 0.801 | 1.051 | $57.6 \pm 2.6$ | $57.3 \pm 3.6$ | $57.5 \pm 2.1 \pm 2.5$ |
| 4.3774 | 522.7 | $476.4 \pm 22.5$ | 44.10 | $349.1 \pm 21.5$ | 40.79 | 0.820 | 1.051 | $58.0 \pm 2.7$ | $62.1 \pm 3.8$ | $59.4 \pm 2.2 \pm 2.6$ |
| 4.3874 | 55.6 | $57.4 \pm 7.8$ | 44.73 | $28.5 \pm 6.8$ | 41.43 | 0.836 | 1.051 | $63.5 \pm 8.6$ | $46.0 \pm 11.0$ | $55.7 \pm 6.4 \pm 2.4$ |
| 4.3964 | 507.8 | $462.9 \pm 22.3$ | 43.46 | $311.5 \pm 20.4$ | 39.96 | 0.855 | 1.051 | $56.4 \pm 2.7$ | $55.8 \pm 3.7$ | $56.2 \pm 2.2 \pm 2.6$ |
| 4.4156 | 1090.7 | $787.3 \pm 29.1$ | 41.88 | $518.2 \pm 27.3$ | 39.58 | 0.908 | 1.052 | $43.6 \pm 1.6$ | $41.1 \pm 2.2$ | $42.7 \pm 1.3 \pm 2.0$ |
| 4.4362 | 569.9 | $326.4 \pm 19.0$ | 39.29 | $214.0 \pm 18.6$ | 36.66 | 0.985 | 1.054 | $33.9 \pm 2.0$ | $32.2 \pm 2.8$ | $33.4 \pm 1.6 \pm 1.5$ |
| 4.4671 | 111.1 | $19.3^{+5.4}_{-4.6}$ | 35.06 | $8.3^{+5.7}_{-4.6}$ | 32.76 | 1.140 | 1.055 | $10.0^{+2.8}_{-2.4}$ | $6.2^{+4.3}_{-3.4}$ | $9.0^{+2.3}_{-2.0} \pm 0.4$ |
| 4.5271 | 112.1 | $14.9^{+4.9}_{-4.0}$ | 27.16 | $10.8^{+5.1}_{-4.3}$ | 22.55 | 1.490 | 1.054 | $7.5^{+2.5}_{-2.0}$ | $8.9^{+4.2}_{-3.5}$ | $7.9^{+2.1}_{-1.8} \pm 0.3$ |
| 4.5745 | 48.93 | $4.5^{+3.1}_{-2.3}$ | 31.34 | $7.0^{+3.9}_{-3.1}$ | 26.10 | 1.210 | 1.054 | $5.5^{+3.9}_{-2.8}$ | $14.0^{+7.8}_{-6.2}$ | $8.1^{+3.6}_{-3.0} \pm 0.4$ |
| 4.5995 | 586.9 | $106.5 \pm 11.1$ | 37.40 | $97.1 \pm 12.1$ | 31.66 | 0.987 | 1.055 | $11.3 \pm 1.2$ | $16.4 \pm 2.0$ | $12.9 \pm 1.0 \pm 0.6$ |
| 4.6121 | 102.5 | $22.6 \pm 5.2$ | 39.28 | $16.2 \pm 5.1$ | 34.54 | 0.923 | 1.055 | $13.9 \pm 3.2$ | $15.4 \pm 4.8$ | $14.4 \pm 2.7 \pm 0.6$ |
| 4.6279 | 511.1 | $161.0 \pm 13.4$ | 41.21 | $108.6 \pm 12.6$ | 36.41 | 0.884 | 1.054 | $19.8 \pm 1.6$ | $20.5 \pm 2.4$ | $20.0 \pm 1.4 \pm 0.9$ |
| 4.6409 | 541.4 | $194.3 \pm 14.7$ | 42.25 | $118.9 \pm 13.4$ | 37.69 | 0.877 | 1.054 | $22.2 \pm 1.7$ | $20.6 \pm 2.3$ | $21.7 \pm 1.4 \pm 1.0$ |
| 4.6613 | 523.6 | $204.0 \pm 15.1$ | 41.64 | $139.7 \pm 14.9$ | 38.10 | 0.898 | 1.054 | $23.9 \pm 1.8$ | $24.2 \pm 2.6$ | $24.0 \pm 1.5 \pm 1.1$ |
| 4.6812 | 1643.4 | $590.1 \pm 25.6$ | 40.18 | $400.6 \pm 25.3$ | 36.31 | 0.939 | 1.054 | $22.0 \pm 1.0$ | $22.3 \pm 1.4$ | $22.1 \pm 0.8 \pm 1.0$ |
| 4.6984 | 526.2 | $166.8 \pm 13.7$ | 39.16 | $114.8 \pm 13.6$ | 35.66 | 0.980 | 1.055 | $18.9 \pm 1.6$ | $19.3 \pm 2.3$ | $18.9 \pm 1.3 \pm 0.8$ |

Mode II. The $J/\psi$ is reconstructed via the decay modes $J/\psi \to l^+l^-(l = e/\mu)$.

## II. THE BESIII EXPERIMENT AND THE DATA SETS

The BESIII detector [19] records symmetric $e^+e^-$ collisions provided by the BEPCII storage ring [20], which operates with a peak luminosity of $1 \times 10^{33}$ cm$^{-2}$ s$^{-1}$ in the c.m. energy range from 2.0 to 4.946 GeV. BESIII has collected large data samples in this energy region [21]. The cylindrical core of the BESIII detector covers 93% of the full solid angle and consists of a helium-based multilayer drift chamber (MDC), a plastic scintillator time-of-flight system (TOF), and a CsI(Tl) electromagnetic calorimeter (EMC), which are all enclosed in a superconducting solenoidal magnet providing a 1.0 T magnetic field. The solenoid is supported by an octagonal flux-return yoke with resistive plate counter muon identification modules interleaved with steel. The charged-particle momentum resolution at 1 GeV$/c$ is 0.5%, and the d$E/$d$x$ resolution is 6% for electrons from Bhabha scattering. The EMC measures photon energies with a resolution of 2.5% (5%) at 1 GeV in the barrel (end cap) region. The time resolution in the TOF barrel region is 68 ps, while that in the end cap region is 110 ps. The end cap TOF system was upgraded in 2015 using multigap resistive plate chamber technology, providing a time resolution of 60 ps [22].

Simulated data samples produced with a GEANT4-based [23] Monte Carlo (MC) package, which includes the geometric description of the BESIII detector and the detector response, are used to determine detection efficiencies and to estimate background contributions. The simulation models beam energy spread and ISR effects in the $e^+e^-$ annihilations with the generator KKMC [24]. The signal process $e^+e^- \to \pi^+\pi^-\psi(3686)$ is simulated by incorporating the line shape of the cross section from the previous study [16]. The inclusive MC sample includes the production of open charm processes, the ISR production of vector charmonium(-like) states, and the continuum processes incorporated in KKMC [24]. The known decay modes are modeled with EVTGEN [25] using branching fractions taken from the Particle Data Group (PDG) [26], and the remaining unknown charmonium decays are modeled with LUNDCHARM [27]. The QED final-state radiation (FSR) correction from charged final-state particles is incorporated using the PHOTOS package [28].

## III. EVENT SELECTION

The topology of a signal event includes either six charged tracks (Mode I) or four charged tracks and at least two photons (Mode II). The good charged tracks detected in the MDC are required to be within a polar angle ($\theta$) range of $|\cos\theta| < 0.93$, where $\theta$ is defined with respect to the $z$ axis. For charged tracks, the distance of closest approach to the interaction point must be less than 10 cm along the $z$ axis, $|V_z| < 10$ cm, and less than 1 cm in the transverse plane, $|V_{xy}| < 1$ cm. Photon candidates are identified using showers in the EMC. The deposited energy of each shower must be more than 25 MeV in the barrel region ($|\cos\theta| < 0.80$) and more than 50 MeV in the end cap region ($0.86 < |\cos\theta| < 0.92$). To exclude showers that originate from charged tracks, the angle between the position of each shower in the EMC and the closest extrapolated charged track must be greater than 10 degrees. To suppress electronic noise and showers unrelated to the event, the difference between the EMC time and the event start time is required to be within (0, 700) ns.

To improve the efficiency, candidate events from Mode I are required to have five charged tracks with $\pm 1$ net charge, allowing for one missed charged pion, or six charged tracks with zero net charge. Candidate events from Mode II are required to have four charged tracks with zero net charge and at least two good photon candidates. For the signal candidates of both Modes, the pions and leptons are kinematically well separated, and thus the charged tracks with momenta above 1.0 GeV$/c$ are assigned to be leptons, and those with momenta below 0.65 (0.80) GeV$/c$ are assigned to be pions for data samples with c.m. energies below (above) 4.465 GeV. Electron and muon separation is performed using the deposited energy ($E$) in the EMC, i.e., both electrons must satisfy $E/p > 0.7$, while both muons $E < 0.45$ GeV, where $p$ is the momentum of charged tracks. Signal candidates are required to have a lepton pair of the same flavor and opposite charge. A vertex fit is applied to charged tracks from Mode I and Mode II, respectively. The results measured with data samples from $J/\psi \to e^+e^-$ and $J/\psi \to \mu^+\mu^-$ agree well with each other.

In Mode I, a four-constraint (4C) kinematic fit imposing energy-momentum conservation under the hypothesis $e^+e^- \to \pi^+\pi^-\pi^+\pi^-l^+l^-$ is implemented for the candidate events with six charged tracks, while a one-constraint (1C) kinematic fit constraining the missing mass to the pion mass is applied for those events with five charged tracks. The corresponding $\chi^2$ is required to satisfy $\chi^2_{4C} < 60$ and $\chi^2_{1C} < 15$, respectively. Afterwards, a clear $J/\psi$ signal is observed as shown in Fig. 1(a). $M(l^+l^-)$ is calculated with the four-momentum of the lepton pair after the 4C (1C) kinematic fit. Candidate events are further required to satisfy $3.05 < M(l^+l^-) < 3.15$ GeV$/c^2$, shown as the region between the close red arrows in Fig. 1(a). To improve the mass resolution of $\psi(3686)$, $M(l^+l^-)$ is constrained to the PDG [26] value of the $J/\psi$ mass, resulting in a kinematic fit with either five constraints (5C) or two constraints (2C). The events with $M(l^+l^-)$ in the signal region are subjected to these 5C or 2C fits and the resulting kinematic variables are used in the subsequent analysis. Finally, the $\psi(3686)$ signal is extracted from the invariant mass of the $l^+l^-\pi^+\pi^-$ system, $M(l^+l^-\pi^+\pi^-)$, where there are four $\pi^+\pi^-$ combinations in an event. The one with $M(l^+l^-\pi^+\pi^-)$ closest to the $\psi(3686)$ nominal mass [26] is selected.

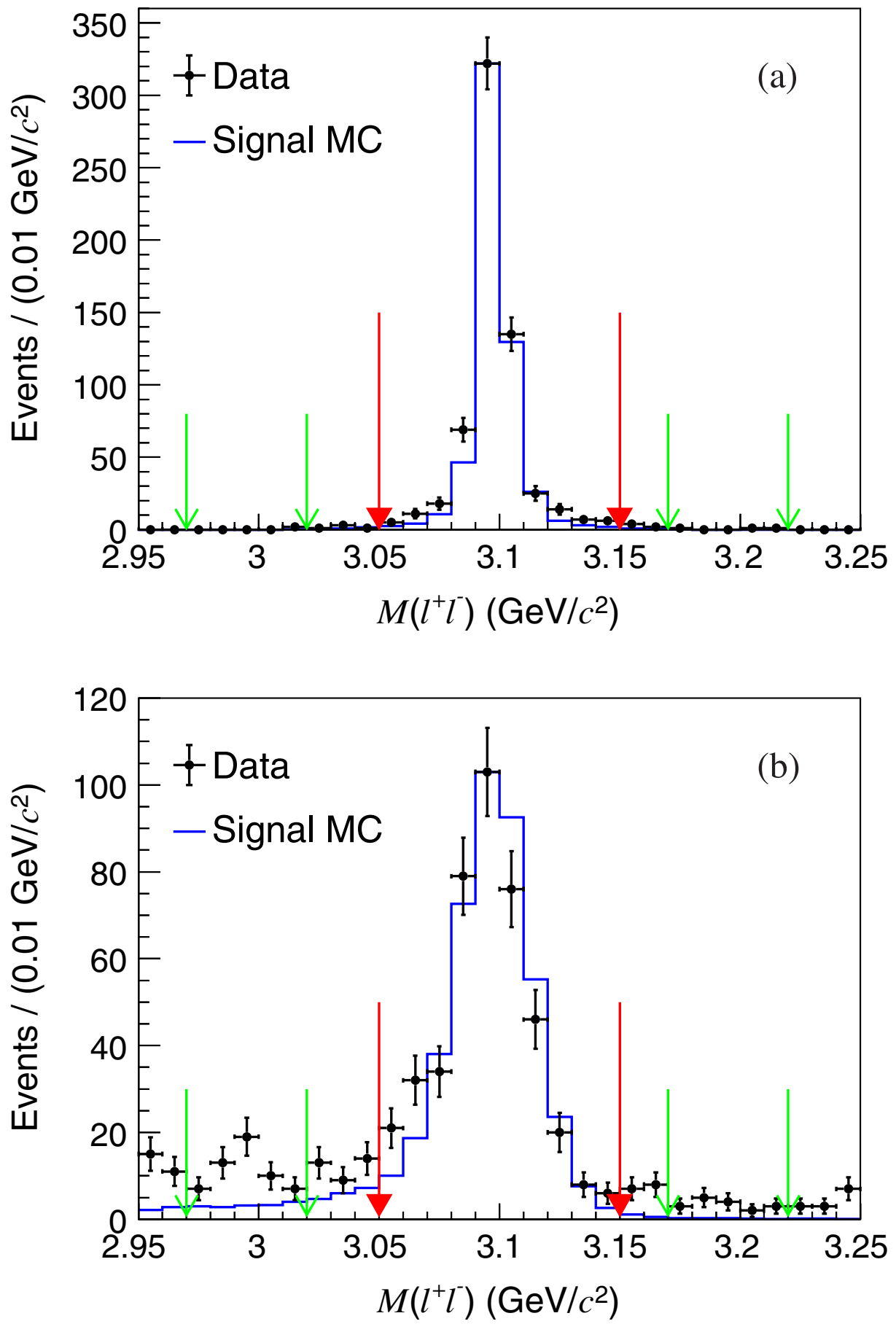

FIG. 1. Distributions of $M(l^+l^-)$ for Mode I (a) after the 4C (1C) kinematic fit and Mode II (b) after applying the selection criteria for the data sample with c.m. energy at 4.3583 GeV. Dots with error bars are data, the blue curve is for signal MC sample, the closed red arrows indicate the signal regions, and the open green arrows indicate the sideband regions.

In Mode II, the number of photons in the final state is not fixed due to the fact that several $\psi(3686)$ decay modes are being included, and no kinematic fit is performed. The $\pi^+\pi^-$ opening angle is required to satisfy $\cos\theta_{\pi^+\pi^-} < 0.9$ in order to remove radiative Bhabha and dimuon background events in which the radiative photon converts into an $e^+e^-$ pair and is misidentified as a $\pi^+\pi^-$ pair. To remove the background due to $e^+e^- \to \pi^+\pi^-J/\psi$, the recoiling mass of the $\pi^+\pi^-l^+l^-$ system $M^{\rm rec}(\pi^+\pi^-l^+l^-)$, which corresponds to the invariant mass of the system of all neutral particles in the final state, is used. Based on exclusive MC simulation, the veto is set at 3 times the width, and events with $|M^{\rm rec}(\pi^+\pi^- l^+ l^-)| > 63\ {\rm MeV}/c^2$ are accepted. To eliminate the background due to $e^+e^- \to {\rm neutrals} + \psi(3686)$ with the subsequent decay $\psi(3686) \to \pi^+\pi^- J/\psi$, the $\psi(3686)$ mass with a $J/\psi$ resolution correction $M^{\rm corr}(\psi(3686)) = M(\pi^+\pi^- l^+ l^-) - M(l^+l^-) + M(J/\psi)$ is required to satisfy $|M^{\rm corr}(\psi(3686)) - M(\psi(3686))| > 8\ {\rm MeV}/c^2$, where $M(\psi(3686))$ is the $\psi(3686)$ mass from the PDG [26]. A requirement of $|M(\gamma\gamma\pi^+\pi^-) - M(\eta)| > 50\ {\rm MeV}/c^2$ is used to eliminate background events from $e^+e^- \to \eta J/\psi$ with a subsequent decay $\eta \to \pi^+\pi^-\pi^0$, where $\gamma\gamma$ are the two photons with the largest energy and $M(\eta)$ is the nominal $\eta$ mass from the PDG [26]. The $J/\psi$ signal is extracted from the $M(l^+l^-)$ distribution by requiring $3.05 < M(l^+l^-) < 3.15\ {\rm GeV}/c^2$ as indicated in Fig. 1(b). From the study of data sets with $\sqrt{s} = 4.1784$, 4.2263, 4.2580, 4.4156 GeV, the small peak around $3.0\ {\rm GeV}/c^2$ in Fig. 1(b) is determined to be due to statistical fluctuation. Finally, the $\psi(3686)$ signal is extracted from the invariant mass recoiling against the $\pi^+\pi^-$ system, $M^{\rm rec}(\pi^+\pi^-)$.

After applying all of the event selection criteria mentioned above, the distributions of $M(l^+l^-\pi^+\pi^-)$ for Mode I and $M^{\rm rec}(\pi^+\pi^-)$ for Mode II are shown in Fig. 2 for the data sample with a c.m. energy at 4.3583 GeV. The $\psi(3686)$ signals are clearly visible with small background contributions. Potential backgrounds without a $J/\psi$ are studied using events in the $J/\psi$ sideband region, where the $J/\psi$ sideband regions are defined as $2.97 < M(l^+l^-) < 3.02\ {\rm GeV}/c^2$ and $3.17 < M(l^+l^-) < 3.22\ {\rm GeV}/c^2$, as shown by the green arrows in Fig. 1. In Mode I, the 5C or 2C kinematic fit is not performed for the sideband events. The corresponding distributions of $J/\psi$ sideband events are shown in Fig. 2, where no peak around the $\psi(3686)$ signal is observed. Potential background events including a $J/\psi$ in the final state are studied using inclusive as well as dedicated exclusive MC samples, and no peak around the $\psi(3686)$ signal is found.

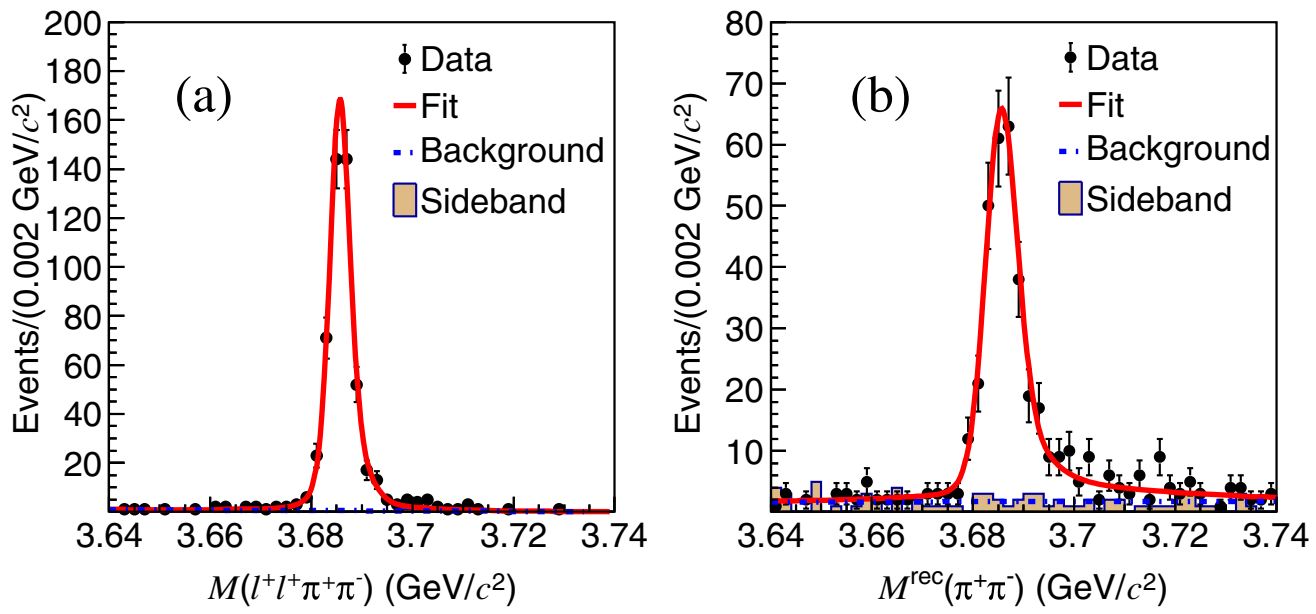

FIG. 2. Distributions of $M(l^+l^-\pi^+\pi^-)$ for Mode I (a) and $M^{\rm rec}(\pi^+\pi^-)$ for Mode II (b) for data with a c.m. energy of 4.3583 GeV. Dots with error bars are data, the shaded histograms represent the background events in the $J/\psi$ sideband regions, the solid curves are the fit results, and the dashed curves are the background contributions from the fit.

## IV. EXTRACTION OF THE BORN CROSS SECTION

The Born cross section of the process is determined by

$$\sigma^B = \frac{N_{\rm obs}}{\mathcal{L}_{\rm int}(1+\delta)\frac{1}{|1-\Pi|^2}\epsilon\mathcal{B}}, \tag{1}$$

where $N_{\rm obs}$ is the observed signal yield, $\mathcal{L}_{\rm int}$ is the integrated luminosity, $(1+\delta)$ is the ISR correction factor obtained from the KKMC generator, $\frac{1}{|1-\Pi|^2}$ is the vacuum polarization factor from QED calculations [29], $\mathcal{B}$ is the product of the branching fractions of the intermediate states, which is 4.14% for Mode I and 3.06% for Mode II, and $\epsilon$ is the detection efficiency estimated by the MC simulation. To consider the effects from the intermediate states and the angular distribution in the detection efficiency, a simple partial wave analysis (PWA) is performed to the data sets with total number of events of Mode I and Mode II larger than 100. The contributions of the $Z_c(3900)$, $Z_c(4020)$, $f_0(500)$ and $f_0(980)$ are considered for data sets with $\sqrt{s} < 4.5995$ GeV and only the contributions from $f_0(500)$ and $f_0(980)$ are considered for data sets with $\sqrt{s} \geq 4.5995$ GeV. The PWA results can describe the data well and the resulting amplitudes are used to generate the signal MC samples. For the data sets with low statistics, the PWA results of the closest high-statistics data sets are used.

The measured cross section for each c.m. energy is obtained by performing simultaneous unbinned maximum likelihood fits to the corresponding $M(l^+l^-\pi^+\pi^-)$ and $M^{\rm rec}(\pi^+\pi^-)$ spectra, where the two decay modes share the same cross section values. The signal shape is described with a MC-simulated shape convolved with a Gaussian function. The Gaussian function represents the mass resolution difference between data and MC simulation, and its parameters are allowed to float during the fit procedure. The background shape is a linear function. The fit curves for the data sample with c.m. energy 4.3583 GeV are shown in Fig. 2. The signal yields for Mode I and Mode II are also given by individual fits. The obtained cross section results for the two $\psi(3686)$ decay modes are consistent with each other. The cross section results and signal yields are summarized in Table I for the different c.m. energies. A comparison shown in Fig. 3 validates the consistency of cross section results among BESIII [16], Belle [14] and *BABAR* [15]. The BESIII results have the best precision.

Several sources of systematic uncertainties are considered in the measurement of the Born cross section and they can be classified into two categories: the shared uncertainties between the two $\psi(3686)$ decay modes and the individual ones. The uncertainties for the combined results of both $\psi(3686)$ decay modes are obtained by the following approaches: the shared uncertainties among the two $\psi(3686)$ decay modes, which are listed with the symbol "†" in Table II, are directly propagated to the combined results, while for the individual ones, the simultaneous fit is repeated by varying the corresponding values in Eq. (1) by one standard deviation, and the largest changes with respect to the nominal results are taken as the systematic uncertainties.

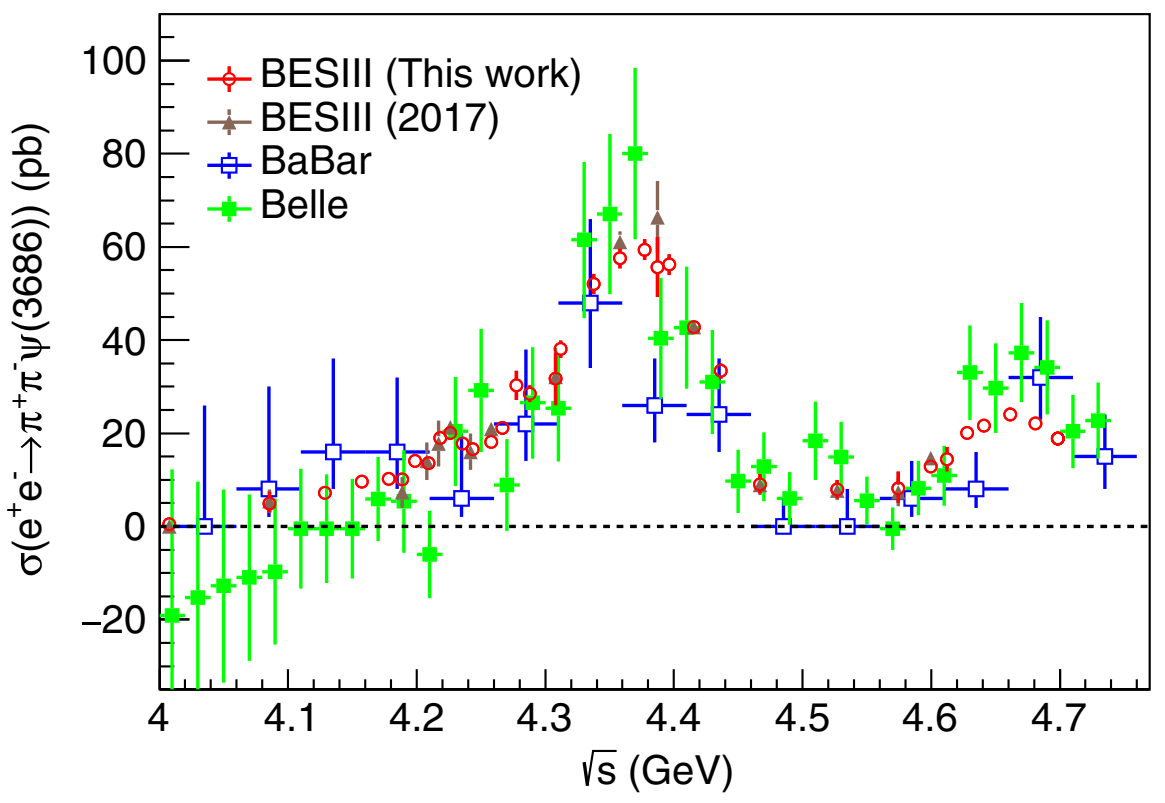

FIG. 3. Comparison of the cross section among this work, previous work of BESIII [16], Belle [14] and *BABAR* [15]. The open circles represent the results of this work. The full triangles are from previous BESIII work. The open squares are from *BABAR* work. The full squares are from Belle work.

The shared uncertainties include those related to the luminosity, the vacuum polarization, the radiative corrections, the lepton tracking efficiency and the residual sources. The uncertainty of the integrated luminosity is 1% as obtained by analysing the large-angle Bhabha scattering events [30]. The uncertainty of the vacuum polarization factor is taken as 0.5% from the QED calculation in Ref. [31]. In Eq. (1), $(1+\delta)$ and $\epsilon$ depend on the input line shape of the cross section due to the effect of ISR. Thus, the measurement of the cross section is performed iteratively until the last variation of $(1+\delta)\epsilon$ becomes negligible, where the measured cross section from this iteration is used as the input of the next iteration. Consequently, the uncertainty related to the ISR correction is dominated by the uncertainty of the input line shape of the Born cross section. To estimate the corresponding uncertainty, the $(1+\delta)\epsilon$ is evaluated 100 times varying the input cross section line shape parameters within the uncertainties obtained from the nominal result of this analysis. The standard deviation of the obtained $(1+\delta)\epsilon$ distribution is considered as the systematic uncertainty. The parametrization of the input cross section will be discussed in the next section. The precision of the ISR calculation in the KKMC generator, 0.5%, is taken as an additional uncertainty related to the ISR correction. The uncertainty in the tracking efficiency of leptons is 1.0% per track [32]. Therefore, two leptons in the final state result in an uncertainty of 2.0%. The residual uncertainty from other sources, e.g., lepton separation, trigger efficiency and FSR is small and is conservatively estimated to be 1.0%.

TABLE II. Summary of the systematic uncertainties (%) for different data sets. The sources marked with "$*$" are shared systematic uncertainties for different data sets, and the sources marked with "†" are shared systematic uncertainties between the two $\psi(3686)$ decay modes. The "$\cdots$" means that the value is very small and can be neglected.

| Data set | 4180 | 4230 | 4360 | 4420 | 4680 |
|---|---|---|---|---|---|
| Luminosity$^{*\dagger}$ | 1.0 | 1.0 | 1.0 | 1.0 | 1.0 |
| $\frac{1}{\lvert 1-\Pi\rvert^2}$ factor$^{*\dagger}$ | 0.5 | 0.5 | 0.5 | 0.5 | 0.5 |
| Tracking$(l^+l^-)^{*\dagger}$ | 2.0 | 2.0 | 2.0 | 2.0 | 2.0 |
| Branching fraction$^*$ | 1.1 | 1.1 | 1.1 | 1.1 | 1.2 |
| Radiative correction$^\dagger$ | 0.5 | 0.7 | 0.5 | 0.5 | 0.5 |
| Tracking$(\pi^+\pi^-)^*$ | 3.0 | 3.0 | 2.9 | 2.9 | 2.9 |
| Photon detection$^*$ | 0.6 | 0.8 | 0.6 | 0.9 | 0.5 |
| $J/\psi$ mass window | 0.2 | 0.2 | 0.3 | 0.4 | 0.1 |
| Kinematic fit | 1.7 | 1.6 | 1.4 | 1.9 | 1.7 |
| $\cos\theta_{\pi^+\pi^-}$ | 0.1 | 0.2 | $\cdots$ | 0.4 | 0.3 |
| $M^{\rm rec}_{\pi^+\pi^-l^+l^-}$ cut | 0.2 | 0.3 | 0.1 | 0.1 | 0.1 |
| $M^{\rm corr}_{\pi^+\pi^-l^+l^-}$ cut | 0.1 | 0.4 | 0.1 | $\cdots$ | $\cdots$ |
| $M_{\gamma\gamma\pi^+\pi^-}$ cut | 0.1 | 0.1 | 0.2 | $\cdots$ | $\cdots$ |
| Background shape | $\cdots$ | $\cdots$ | 0.3 | $\cdots$ | 0.1 |
| Signal shape | 0.8 | 0.7 | 0.3 | 0.3 | 0.6 |
| Fitting range | 1.1 | 1.9 | 0.3 | 0.7 | 0.3 |
| MC models | 1.1 | 1.0 | 1.0 | 1.0 | 1.1 |
| Others$^{*\dagger}$ | 1.0 | 1.0 | 1.0 | 1.0 | 1.0 |
| Total | 4.8 | 5.1 | 4.5 | 4.7 | 4.6 |

The individual uncertainties include contributions from the branching fractions, the tracking efficiency of pions and photons, the $J/\psi$ mass window, the kinematic fit, the requirement to suppress backgrounds, the fit procedure, and the MC model. The uncertainties of the branching fractions of the $\psi(3686)$ and other intermediate states are taken from the PDG [26]. The uncertainty in the tracking efficiency for pions is 1.0% per track [32]. A weighted value of 3.5% is assigned to Mode I since the yields ratio between the three- and four-pion cases is about 1, and 2.0% for Mode II with two pions in the final state. The uncertainty in the photon detection efficiency is 1.0% per photon [33]. For Mode II with at least two photons, an alternative measurement of the cross sections with a 2.0% efficiency change is performed to determine the relative differences, which are assigned as the uncertainties from the photon detection efficiency. The uncertainty associated with the $J/\psi$ mass requirement is estimated by smearing the $M(l^+l^-)$ distribution of MC samples according to the resolution discrepancy between data and MC, and the resulting uncertainties in detection efficiencies are about 0.1% for Mode I and 0.8% for Mode II. In Mode I, the uncertainty for the kinematic fit is estimated by tuning the helix parameters of charged tracks according to data as in Ref. [34], and the case without the helix parameters correction is taken as the nominal case. In Mode II, the uncertainties associated with the requirements of $\cos\theta_{\pi^+\pi^-}$, $M^{\rm rec}(\pi^+\pi^-l^+l^-)$, $M^{\rm corr}(\psi(3686))$ and $M(\gamma\gamma\pi^+\pi^-)$ are evaluated by comparison to the alternative conditions $\cos\theta_{\pi^+\pi^-} < 0.8$, $|M^{\rm rec}(\pi^+\pi^-l^+l^-)| > 105\ {\rm MeV}/c^2$, $|M^{\rm corr}(\psi(3686)) - M(\psi(3686))| > 10\ {\rm MeV}/c^2$ and $|M(\gamma\gamma\pi^+\pi^-) - M(\eta)| > 60\ {\rm MeV}/c^2$, respectively. The uncertainties related to the fit procedure are investigated by changing the fit range, replacing the linear function by a quadratic function for the background description and by varying the width of the convolved Gaussian function for the signal shape. The selection efficiency is obtained with the signal MC sample generated according to the PWA results. To estimate the corresponding uncertainty of the MC model, 100 sets of signal MC samples are generated to obtain the detection efficiency distribution, and the resulting standard deviation is taken as the contribution to the systematic uncertainty. In each set, the MC sample is generated by varying all the PWA parameters randomly according to a multivariate Gaussian function, where the mean and width are the nominal value and error of the parameters with correlation considered.

The uncertainties for the combined results of the data samples with large statistics are summarized in Table II. For those data samples with low statistics, the uncertainties are set as the values of the closest data sets in Table II. Assuming all sources of systematic uncertainties to be independent, the total uncertainties are obtained by adding the individual values in quadrature and are found to be in the range of 4.5% to 5.1%.

## V. FIT TO THE CROSS SECTION

To study possible $Y$ states in the process $e^+e^- \to \pi^+\pi^-\psi(3686)$, a binned $\chi^2$ fit is performed to the dressed cross sections $\sigma^{\rm dressed} = \sigma^B \cdot (\frac{1}{|1-\Pi|^2})$. The $\chi^2$ is constructed according to the method in Ref. [35] by incorporating both statistical and systematical uncertainties and considering both the correlated and uncorrelated terms according to the formula

$$\chi^2 = \Delta X^T M^{-1} \Delta X, \quad (2)$$

where $\Delta X$ is the difference between the measured dressed cross section and the expected value calculated by the probability density function (PDF) for each signal c.m. energy. $M$ is the covariance matrix of elements

$$M_{ij} = \begin{cases} (\Delta_i^{\rm sys})^2 + (\Delta_i^{\rm stat})^2 & i = j, \\ (\sigma_i \cdot \epsilon_s) \cdot (\sigma_j \cdot \epsilon_s) & i \neq j, \end{cases} \quad (3)$$

where the index $i(j)$ represents the $i(j)$th data set, $\Delta_i^{\rm stat}$ is the asymmetric statistic uncertainty for the $i$th data set and the value is chosen following the strategy in Ref. [10], $\Delta_i^{\rm sys}$ is the total systematic uncertainty, and $\sigma_i(\sigma_j)$ is the measured cross section. $\epsilon_s$ is the relative correlated systematic uncertainty, which is the quadratic sum of terms marked with "*" in Table II. Because some correlated uncertainties are different for different data sets, to be conservative, the largest value is used for the line shape fit.

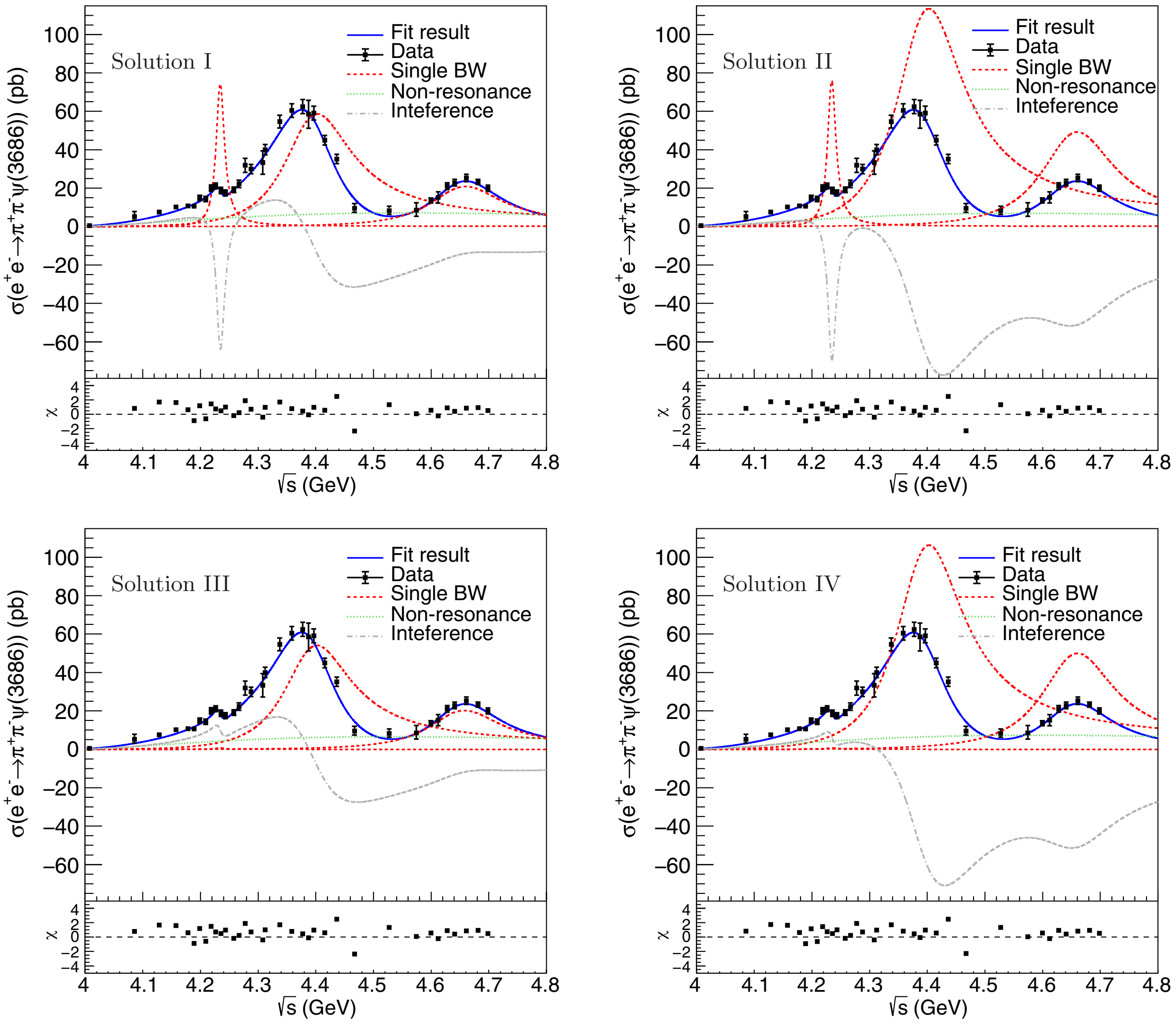

FIG. 4. The dressed cross section fit results of the process $e^+e^- \to \pi^+\pi^-\psi(3686)$ corresponding to the four solutions in Table III. The black dots with error bars are the measured dressed cross section, the blue solid curves are the best-fit results, the red dashed lines represent individual resonant structures, the green dotted lines show the continuous component, and the gray dot-dashed lines are the sum of all interference terms. The bottom panel in each plot is the $\chi$ distribution.

The PDF used in the fit is parametrized as a coherent sum of the possible resonant components together with a phase space component for the continuous contribution, i.e.,

$$\sigma^{\rm dressed}(\sqrt{s}) = \left| \sum_k e^{i\phi_k} \cdot BW_k(s) + e^{i\phi_{\rm cont}} \cdot \psi_{\rm cont} \right|^2, \quad (4)$$

where $BW$ is the Breit-Wigner function with the three-body phase space (PHSP) factor

$$BW_k(s) = \frac{M_k}{\sqrt{s}} \frac{\sqrt{12\pi\Gamma_k^{\rm tot}\Gamma_k^{ee}B_k}}{s - M_k^2 + iM_k\Gamma_k^{\rm tot}} \sqrt{\frac{\Phi(\sqrt{s})}{\Phi(M_k)}}, \quad (5)$$

representing the possible resonant structures $Y(4220)$, $Y(4390)$ and $Y(4660)$. $\phi_k$ and $\phi_{\rm cont}$ are the corresponding phases relative to that of the $Y(4390)$. The continuous part is parametrized as

$$\psi_{\rm cont} = \frac{a}{(\sqrt{s})^n} \sqrt{\Phi(\sqrt{s})}. \quad (6)$$

In Eq. (5), $M_k$, $\Gamma_k^{\rm tot}$, $\Gamma_k^{ee}$ and $B_k$ are the parameters representing mass, full width, partial width coupling to $e^+e^-$, and the branching fraction to the $\pi^+\pi^-\psi(3686)$ final state. $\Phi(\sqrt{s})$ is the standard three-body PHSP factor [26]. In Eq. (6), $a$ and $n$ are free parameters. A nominal fit including $Y(4220)$, $Y(4390)$ and $Y(4660)$ as well as the continuous components is shown in Fig. 4, and the results are summarized in Table III. The fit yields four solutions with equal fit quality $\chi^2/{\rm d.o.f.} = 31.0/20$, where d.o.f. is the number of degrees of freedom. The fit without the continuous component is also tested, with the resulting fit quality $\chi^2/{\rm d.o.f.} = 56.3/23$, which indicates the necessity of a continuous component.

An alternative fit is carried out by parametrizing the continuous component with an exponential function as used in the analysis of $e^+e^- \to \pi^+\pi^- J/\psi$ [7,36],

$$\sqrt{\sigma_{NY}} = \sqrt{\Phi(\sqrt{s})e^{-p_0 u} p_1}, \quad (7)$$

where $p_0$, $p_1$ are floating parameters and $u = \sqrt{s} - (2M(\pi^\pm) + M(\psi(3686)))$. The fit quality is $\chi^2/{\rm d.o.f.} = 31.1/20$ and the resonant parameters are almost the same as the nominal fit results. In this case, the fit results are listed in Table IV.

TABLE III. Results of the fit to the $e^+e^- \to \pi^+\pi^-\psi(3686)$ cross section for the case when the continuous part is described by $\psi_{\rm cont}$ in Eq. (6). The uncertainties involve statistical and systematic ones propagated from the cross section measurement in Table I.

| Parameters | Solution I | Solution II | Solution III | Solution IV |
|---|---|---|---|---|
| M($Y$4220) (MeV/$c^2$) | | $4234.4 \pm 3.2$ | | |
| $\Gamma^{\rm tot}$($Y$4220) (MeV) | | $17.6 \pm 8.1$ | | |
| $B\Gamma^{ee}$($Y$4220) (eV) | $1.59 \pm 0.75$ | $1.63 \pm 0.78$ | $0.02 \pm 0.01$ | $0.02 \pm 0.01$ |
| M($Y$4390) (MeV/$c^2$) | | $4390.3 \pm 6.0$ | | |
| $\Gamma^{\rm tot}$($Y$4390) (MeV) | | $143.3 \pm 10.0$ | | |
| $B\Gamma^{ee}$($Y$4390) (eV) | $10.70 \pm 4.13$ | $20.72 \pm 2.46$ | $9.86 \pm 4.11$ | $19.44 \pm 2.04$ |
| M($Y$4660) (MeV/$c^2$) | | $4651.0 \pm 37.8$ | | |
| $\Gamma^{\rm tot}$($Y$4660) (MeV) | | $155.4 \pm 24.8$ | | |
| $B\Gamma^{ee}$($Y$4660) (eV) | $4.72 \pm 3.79$ | $11.15 \pm 3.23$ | $4.66 \pm 4.20$ | $11.28 \pm 3.25$ |
| $\phi_{Y(4220)}$ (rad) | $1.68 \pm 0.04$ | $1.39 \pm 0.06$ | $6.24 \pm 0.05$ | $5.95 \pm 0.04$ |
| $\phi_{Y(4660)}$ (rad) | $6.07 \pm 0.03$ | $4.77 \pm 0.04$ | $6.03 \pm 0.06$ | $4.77 \pm 0.03$ |
| $\phi_{\rm cont}$ (rad) | $3.14 \pm 0.79$ | $2.58 \pm 0.13$ | $2.99 \pm 0.92$ | $2.40 \pm 0.08$ |
| $a(\times 10^5)$ | $4.81 \pm 35.83$ | $5.28 \pm 33.23$ | $5.13 \pm 26.55$ | $3.48 \pm 24.16$ |
| $n$ | $8.65 \pm 3.66$ | $8.72 \pm 3.40$ | $8.69 \pm 3.09$ | $8.43 \pm 3.53$ |

TABLE IV. Results of the fit to the $e^+e^- \to \pi^+\pi^-\psi(3686)$ cross section when the continuous part is described by $\sigma_{NY}$ in Eq. (7). The uncertainties involve statistical and systematic ones propagated from the cross section measurement in Table I.

| Parameters | Solution I | Solution II | Solution III | Solution IV |
|---|---|---|---|---|
| M($Y$4220) (MeV/$c^2$) | $4234.2 \pm 3.5$ | | | |
| $\Gamma^{\rm tot}$($Y$4220) (MeV) | $18.0 \pm 8.8$ | | | |
| $B\Gamma^{ee}$($Y$4220) (eV) | $1.63 \pm 0.82$ | $1.64 \pm 0.83$ | $0.02 \pm 0.01$ | $0.02 \pm 0.01$ |
| M($Y$4390) (MeV/$c^2$) | $4390.9 \pm 7.4$ | | | |
| $\Gamma^{\rm tot}$($Y$4390) (MeV) | $143.6 \pm 11.3$ | | | |
| $B\Gamma^{ee}$($Y$4390) (eV) | $10.64 \pm 4.36$ | $19.73 \pm 5.57$ | $9.79 \pm 3.46$ | $19.12 \pm 2.01$ |
| M($Y$4660) (MeV/$c^2$) | $4652.5 \pm 41.0$ | | | |
| $\Gamma^{\rm tot}$($Y$4660) (MeV) | $154.9 \pm 25.3$ | | | |
| $B\Gamma^{ee}$($Y$4660) (eV) | $4.75 \pm 4.28$ | $10.21 \pm 5.21$ | $4.69 \pm 3.59$ | $10.58 \pm 3.78$ |
| $\phi_{Y(4220)}$ (rad) | $1.68 \pm 0.04$ | $1.44 \pm 0.05$ | $6.27 \pm 0.05$ | $6.03 \pm 0.04$ |
| $\phi_{Y(4660)}$ (rad) | $6.07 \pm 0.04$ | $4.65 \pm 0.04$ | $6.03 \pm 0.04$ | $4.71 \pm 0.04$ |
| $\phi_{\sigma_{\rm NY}}$ (rad) | $3.14 \pm 0.83$ | $2.70 \pm 0.55$ | $2.99 \pm 0.67$ | $2.45 \pm 0.13$ |
| $p_0(\times 10^5)$ | $3.80 \pm 2.49$ | $4.04 \pm 2.66$ | $3.79 \pm 2.44$ | $3.70 \pm 1.92$ |
| $p_1$ | $9.47 \pm 5.46$ | $9.79 \pm 5.60$ | $9.50 \pm 5.37$ | $9.09 \pm 3.82$ |

Following Ref. [37], the continuous component is also described with a Fano-like interference function,

$$\sigma_{\rm fano}(u) = g_1 u^2 e^{-g_2 u^2}, \quad (8)$$

where $g_1$ and $g_2$ are free parameters and the $u$ is the same as that in Eq. (7). The resulting fit quality is $\chi^2/{\rm d.o.f.} = 40.0/20$, so this fit method is not accepted.

To examine the significance of the three $Y$ states, the fits are repeated using only two of the three $Y$ states. The significances for the $Y(4220)$, $Y(4390)$, and $Y(4660)$ are $4.0\sigma$, $16.1\sigma$, and $8.1\sigma$, respectively. Also, a fit is performed by fixing the resonant parameters of the $Y(4220)$ to be those of the $\psi(4160)$ from the PDG [26] and the resulting fit quality is $\chi^2/{\rm d.o.f.} = 55.5/22$. The fit quality indicates that the contribution from the $Y(4220)$ cannot be replaced by the $\psi(4160)$. It is worth noting that the width of the $Y(4220)$ in the nominal results is much smaller than that of the previous BESIII analysis [16], which was $\Gamma^{\rm tot}$= 80.1 MeV. If the width of the $Y(4220)$ is fixed to this value, the fit yields $\chi^2/{\rm d.o.f.} = 54.3/21$. To examine the existence of extra resonances, fits including an additional resonance with fixed mass and width taken from the PDG [26], e.g., $\psi(4160)$ and $\psi(4415)$, or floating mass and width are performed. However, none of the extra resonances are observed with a significance larger than $3\sigma$.

TABLE V. Summary of the systematic uncertainties of the resonance parameters in the fit to the cross section of $e^+e^- \to \pi^+\pi^-\psi(3686)$. The "$\cdots$" indicates that the value is very small and can be neglected.

| Source | Solution | Energy scale | Energy spread | Model | Total |
|---|---|---|---|---|---|
| M($Y$4220) (MeV/$c^2$) | - | 0.0 | 0.0 | 0.2 | 0.2 |
| $\Gamma^{\rm tot}$($Y$4220) (MeV) | - | 0.6 | 0.6 | 0.4 | 0.9 |
| M($Y$4390) (MeV/$c^2$) | - | 0.3 | 0.3 | 0.6 | 0.7 |
| $\Gamma^{\rm tot}$($Y$4390) (MeV) | - | 0.4 | 0.4 | 0.3 | 0.5 |
| M($Y$4660) (MeV/$c^2$) | - | 1.0 | 1.0 | 1.5 | 2.1 |
| $\Gamma^{\rm tot}$($Y$4660) (MeV) | - | 0.3 | 0.3 | 0.7 | 0.8 |
| $B\Gamma^{ee}$($Y$4220) (eV) | I | 0.04 | 0.06 | 0.06 | 0.09 |
| | II | 0.01 | 0.04 | 0.04 | 0.06 |
| | III | $\cdots$ | $\cdots$ | $\cdots$ | $\cdots$ |
| | IV | $\cdots$ | $\cdots$ | $\cdots$ | $\cdots$ |
| $B\Gamma^{ee}$($Y$4390) (eV) | I | 0.06 | 0.15 | 0.15 | 0.22 |
| | II | 0.99 | 0.06 | 0.06 | 1.00 |
| | III | 0.07 | 0.10 | 0.01 | 0.12 |
| | IV | 0.32 | 0.02 | 0.03 | 0.32 |
| $B\Gamma^{ee}$($Y$4660) (eV) | I | 0.03 | 0.15 | 0.15 | 0.21 |
| | II | 0.94 | 0.02 | 0.02 | 0.94 |
| | III | 0.03 | 0.15 | 0.10 | 0.18 |
| | IV | 0.70 | 0.10 | 0.03 | 0.71 |

The systematic uncertainties for the resonant parameters in the cross section fit are discussed below. The uncertainty of the c.m. energy is 0.8 MeV, which is obtained from a measurement of dimuon events [38]. To estimate its effects on the cross section fit, the uncertainty of the c.m. energy is included in the construction of the $\chi^2$, and the fit is repeated. The resulting changes with respect to the nominal values are taken as the systematic uncertainties. The c.m. energy spread of BEPCII is 1.6 MeV, which is determined by the Beam Energy Measurement System [39]. Thus, the fit is repeated by convolving a Gaussian function with a width of 1.6 MeV to the PDF of the resonant structures, and the resulting differences are taken as the systematic uncertainties. The uncertainties associated with the PDF modeling are considered to be the differences of the results between the alternative fit with $\sqrt{\sigma_{NY}}$ in Eq. (7) for continuous components and the nominal fit. Table V summarizes the systematic uncertainties for the mass and total width of the resonances $Y(4220)$, $Y(4390)$ and $Y(4660)$, where the total uncertainties are the quadratic sums of the individual values.

## VI. SUMMARY

Using data with an integrated luminosity of 20.1 fb$^{-1}$ collected by the BESIII detector operating at the BEPCII collider, the cross section of the process $e^+e^- \to \pi^+\pi^-\psi(3686)$ is measured at c.m. energies between 4.0076 and 4.6984 GeV. The fit results confirm the existence of the $Y(4220)$, $Y(4390)$ and $Y(4660)$ as well as the contribution of a continuous component. The nominal fit yields a $\chi^2/{\rm d.o.f.} = 31.0/20$ and the masses and widths of the $Y$ states are determined to be $M = 4234.4 \pm 3.2 \pm 0.2$ MeV$/c^2$, $\Gamma = 17.6 \pm 8.1 \pm 0.9$ MeV for the $Y(4220)$, $M = 4390.3 \pm 6.0 \pm 0.7$ MeV$/c^2$, $\Gamma = 143.3 \pm 10.0 \pm 0.5$ MeV for the $Y(4390)$ and $M = 4651.0 \pm 37.8 \pm 2.1$ MeV$/c^2$, $\Gamma = 155.4 \pm 24.8 \pm 0.8$ MeV for the $Y(4660)$, respectively, where the first uncertainties involve statistical and systematic ones propagated from the cross section measurement and the second are systematic from the line shape fit procedure. This is a supersession of the previous BESIII result [16] with improved precision.

A comparison of the masses and widths of the $Y(4220)$ and $Y(4390)$ among different processes investigated at the BESIII experiment is shown in Fig. 5(a), where the parameters of the $Y(4390)$ are consistent among different decay modes within uncertainties. However, the width of the $Y(4220)$ in this analysis is smaller than results from others [7–12] and the previous measurement [16]. The differences of the parameters require more study from not only experiment but also theory. The masses and widths of the $Y(4360)$, $Y(4390)$ and $Y(4660)$ observed in the process $e^+e^- \to \pi^+\pi^-\psi(3686)$ are compared among BESIII, Belle [14] and *BABAR* [15], as shown in Fig. 5(b). The mass of the $Y(4390)$ observed by BESIII is larger than that of the $Y(4360)$ observed by Belle and *BABAR*, while the masses of the $Y(4660)$ are in good agreement among the three experiments. Owing to the large scan data sets collected in BESIII, finer structures are observed in the $Y(4360)$ energy region. From Fig. 3, the $Y(4390)$ observed by BESIII can be considered as the same resonant structure as the $Y(4360)$ observed by Belle and *BABAR* in the same process. The improvement of size and energy sampling on the BESIII data set as well as the interference of finer structures likely account for the mass difference. These results may be further improved in the future by additional energy points and a better understanding of possible intermediate decay modes.

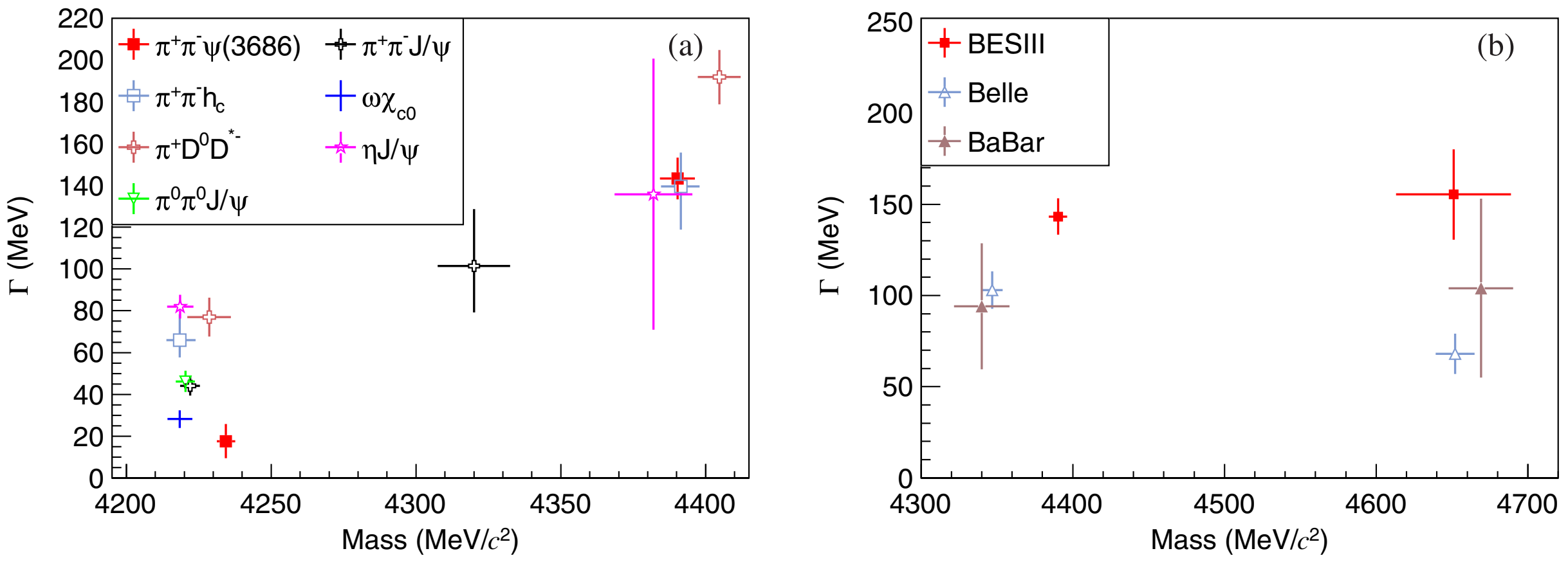

FIG. 5. (a) Masses versus widths of the $Y(4220)$ and $Y(4390)$ obtained from different processes in BESIII [7–12]. (b) Masses versus widths of resonances observed in $e^+e^- \to \pi^+\pi^-\psi(3686)$ from this work and other experiments [14,15].

## ACKNOWLEDGMENTS

The BESIII Collaboration thanks the staff of BEPCII and the IHEP computing center and the supercomputing center of USTC for their strong support. This work is supported in part by National Key R&D Program of China under Contracts Nos. 2020YFA0406300, 2020YFA0406400; National Natural Science Foundation of China (NSFC) under Contracts Nos. 11335008, 11625523, 11635010, 11735014, 11822506, 11835012, 11935015, 11935016, 11935018, 11961141012, 12022510, 12025502, 12035009, 12035013, 12061131003, 11705192, 11950410506, 12061131003; the Chinese Academy of Sciences (CAS) Large-Scale Scientific Facility Program; Joint Large-Scale Scientific Facility Funds of the NSFC and CAS under Contracts Nos. U1732263, U1832207, U1832103, U2032111; CAS Key Research Program of Frontier Sciences under Contract No. QYZDJ-SSW-SLH040; 100 Talents Program of CAS; INPAC and Shanghai Key Laboratory for Particle Physics and Cosmology; ERC under Contract No. 758462; European Union Horizon 2020 research and innovation programme under Contract No. Marie Sklodowska-Curie grant agreement No. 894790; German Research Foundation DFG under Contracts Nos. 443159800, Collaborative Research Center CRC 1044, FOR 2359, FOR 2359, GRK 214; Istituto Nazionale di Fisica Nucleare, Italy; Ministry of Development of Turkey under Contract No. DPT2006K-120470; National Science and Technology fund; Olle Engkvist Foundation under Contract No. 200-0605; STFC (United Kingdom); The Knut and Alice Wallenberg Foundation (Sweden) under Contract No. 2016.0157; The Royal Society, UK under Contracts Nos. DH140054, DH160214; The Swedish Research Council; U.S. Department of Energy under Contracts Nos. DE-FG02-05ER41374, DE-SC-0012069.

---

[1] H. X. Chen, W. Chen, X. Liu, and S. L. Zhu, Phys. Rep. **639**, 1 (2016).
[2] B. Aubert *et al.* (*BABAR* Collaboration), Phys. Rev. Lett. **95**, 142001 (2005).
[3] Q. He *et al.* (CLEO Collaboration), Phys. Rev. D **74**, 091104 (2006).
[4] T. Coan *et al.* (CLEO Collaboration), Phys. Rev. Lett. **96**, 162003 (2006).
[5] C. Yuan *et al.* (Belle Collaboration), Phys. Rev. Lett. **99**, 182004 (2007).
[6] Z. Liu *et al.* (Belle Collaboration), Phys. Rev. Lett. **110**, 252002 (2013); **111**, 019901(E) (2013).
[7] M. Ablikim *et al.* (BESIII Collaboration), Phys. Rev. Lett. **118**, 092001 (2017).
[8] M. Ablikim (BESIII Collaboration), Phys. Rev. D **102**, 012009 (2020).
[9] M. Ablikim *et al.* (BESIII Collaboration), Phys. Rev. D **99**, 091103 (2019).
[10] M. Ablikim *et al.* (BESIII Collaboration), Phys. Rev. Lett. **118**, 092002 (2017).
[11] M. Ablikim *et al.* (BESIII Collaboration), Phys. Rev. Lett. **122**, 102002 (2019).
[12] M. Ablikim *et al.* (BESIII Collaboration), Phys. Rev. D **102**, 031101 (2020).
[13] X. Wang *et al.* (Belle Collaboration), Phys. Rev. Lett. **99**, 142002 (2007).
[14] X. Wang *et al.* (Belle Collaboration), Phys. Rev. D **91**, 112007 (2015).
[15] J. Lees *et al.* (*BABAR* Collaboration), Phys. Rev. D **89**, 111103 (2014).
[16] M. Ablikim *et al.* (BESIII Collaboration), Phys. Rev. D **96**, 032004 (2017); **99**, 019903(E) (2019).
[17] M. Ablikim *et al.* (BESIII Collaboration), Phys. Rev. D **101**, 012008 (2020).
[18] M. Ablikim *et al.* (BESIII Collaboration), Phys. Rev. Lett. **126**, 102001 (2021).
[19] M. Ablikim *et al.* (BESIII Collaboration), Nucl. Instrum. Methods Phys. Res., Sect. A **614**, 345 (2010).
[20] C. Yu *et al.*, in *Proceedings of the 7th International Particle Accelerator Conference* (JACoW, Geneva, 2016), pp. 1014–1018, http://jacow.org/ipac2016/papers/tuya01.pdf.
[21] M. Ablikim *et al.* (BESIII Collaboration), Chin. Phys. C **44**, 040001 (2020).
[22] X. Li *et al.*, Radiat. Detect. Technol. Methods **1**, 13 (2017); Y. X. Guo *et al.*, Radiat. Detect. Technol. Methods **1**, 15 (2017); P. Cao *et al.*, Nucl. Instrum. Methods Phys. Res., Sect. A **953**, 163053 (2020).
[23] S. Agostinelli *et al.* (GEANT4 Collaboration), Nucl. Instrum. Methods Phys. Res., Sect. A **506**, 250 (2003).
[24] S. Jadach, B. F. L. Ward, and Z. Wąs, Phys. Rev. D **63**, 113009 (2001).
[25] D. Lange, Nucl. Instrum. Methods Phys. Res., Sect. A **462**, 152 (2001); R. G. Ping, Chin. Phys. C **32**, 599 (2008).
[26] P. Zyla *et al.* (Particle Data Group Collaboration), Prog. Theor. Exp. Phys. **2020**, 083C01 (2020).
[27] J. C. Chen, G. S. Huang, X. R. Qi, D. H. Zhang, and Y. S. Zhu, Phys. Rev. D **62**, 034003 (2000).
[28] E. Barberio and Z. Wąs, Comput. Phys. Commun. **79**, 291 (1994).
[29] S. Actis *et al.*, Eur. Phys. J. C **66**, 585 (2010).
[30] M. Ablikim *et al.* (BESIII Collaboration), Chin. Phys. C **39**, 093001 (2015).
[31] E. Kuraev and V. S. Fadin, Sov. J. Nucl. Phys. **41**, 466 (1985).

[32] M. Ablikim *et al.* (BESIII Collaboration), Phys. Rev. Lett. **110**, 252001 (2013).
[33] M. Ablikim *et al.* (BESIII Collaboration), Phys. Rev. D **83**, 112005 (2011).
[34] M. Ablikim *et al.* (BESIII Collaboration), Phys. Rev. D **87**, 012002 (2013).
[35] G. D'Agostini, Nucl. Instrum. Methods Phys. Res., Sect. A **346**, 306 (1994).
[36] J. Lees *et al.* (*BABAR* Collaboration), Phys. Rev. D **86**, 051102 (2012).
[37] J. Z. Wang, D. Y. Chen, X. Liu, and T. Matsuki, Phys. Rev. D **99**, 114003 (2019).
[38] M. Ablikim *et al.* (BESIII Collaboration), Chin. Phys. C **40**, 063001 (2016).
[39] E. Abakumova *et al.*, Nucl. Instrum. Methods Phys. Res., Sect. A **659**, 21 (2011).